\documentclass[amsmath,superscriptaddress,showpacs,prb,twocolumn] {revtex4}
\usepackage{bm}
\usepackage{enumerate}
\usepackage{graphicx}
\usepackage[dvips]{epsfig}
\usepackage{epsf}
\usepackage{physics}

\makeatletter
\newcommand{\Rmnum}[1]{\expandafter\@slowromancap\romannumeral #1@}
\makeatletter

\begin{document}
\title{On the magnetotransport of Weyl semimetals due to chiral anomaly}
\author{Vladimir A. Zyuzin}
\affiliation{Department of Physics, University of Florida, P.O. Box 118440, Gainesville, Florida 32611-8440, USA}
\affiliation{Department of Physics and Astronomy and Nebraska Center for Materials and Nanoscience, University of Nebraska, Lincoln, Nebraska 68588, USA}
\begin{abstract}
We study electric field and temperature gradient driven magnetoconductivity of a Weyl semimetal system. 
To analyze the responses, we utilize the kinetic equation with semiclassical equations of motion modified by the Berry curvature and orbital magnetization of the wave-packet. Apart from known positive quadratic magnetoconductivity, we show that due to chiral anomaly, the magnetconductivity can become non-analytic function of the magnetic field, proportional to 3/2 power of the magnetic field at finite temperatures. 
We also show that time-reversal symmetry breaking tilt of the Dirac cones results in linear magnetoconductivity. This is due to one-dimensional chiral anomaly the tilt is responsible for.
\end{abstract}

\maketitle

\textit{Introduction}.
Three dimensional Dirac and Weyl semimetals are materials whose band structure has a linearly touching conduction and valence bands, \cite{Murakami, Ran, Savrasov, BB} the Dirac cones. 
Dirac semimetal is degenerate in electron's right and left chiralities, while the Weyl semimetal has the two chiralities split in energy or momentum. 
Inversion or time-reversal symmetries must be broken to obtain the splitting of chiralities in Dirac semimetal. 

Theoretically, the linear band touching introduces the non-trivial Berry \cite{Berry} curvature in to the description of the fermion dynamics. 
The Berry curvature in this case is an effective magnetic field in ${\bf k} -$ space which is created by a magnetic monopole located at the band touching point. 
For a review on effects of Berry curvature on electronic properties see Ref. [\onlinecite{Niu_rmp}].

Weyl semimetals with broken time-reversal symmetry are characterized by the anomalous Hall effect.\cite{BB,ZB} 
Due to splitting of the Dirac cones, there are chiral edge states on the physical boundaries of the system.\cite{Savrasov,BB} 
Apart from that there is the so-called chiral anomaly of the Dirac fermions - non-conservation of particles with a given chirality in presence of magnetic and electric fields.\cite{Adler, BellJackiw, NN} 

The chiral anomaly results in novel magnetotransport properties which serve as distinctive features of Weyl semimetals.
In Refs. [\onlinecite{SonSpivak,SpivakAndreev,BurkovPRL}] it was shown that the anomaly contributes to the longitudinal quadratic magnetoconductivity. 
In contrast to the Lorentz force driven negative magnetoconductivity, the one due to anomaly was shown to be positive in sign when the fields are collinear.  
This novel magnetic field dependent contribution to the current flows in the direction of the applied magnetic field. 
Signatures of positive magnetoconductivity were experimentally observed in number of materials believed to be Weyl semimetals. 
For example, these are $\mathrm{Cd}_{3}\mathrm{As}_{2}$,\cite{cd3as2_exp1, cd3as2_exp2, cd3as2_exp3} $\mathrm{Na}_{3}\mathrm{Bi}$,\cite{na3bi_exp} $\mathrm{ZrTe}_{5}$,\cite{zrte5_exp1} and TaAs.\cite{taas_exp1, taas_exp2}

In this paper we extend the study of the the nature of the chiral anomaly induced magnetoconductivity in Weyl semimetals. 
Just like in works [\onlinecite{SonSpivak,SpivakAndreev}] we utilize the kinetic equation to describe the transport properties of the system.
The kinetic equation for the wave-packet is supplimented with the equations of motion modified by the Berry curvature and orbital magnetization.
We explicitly focus on the inter-valley scattering processes in collision integral, and show how these processes stabilize steady state in case of chiral anomaly.
For the sake of generality we study electric currents driven by electric field and temperature gradient.
In case of electric field, at small magnetic fields and small temperatures we reproduce the result of Refs. [\onlinecite{SonSpivak,SpivakAndreev}] for the magnetoconductivity.  
However at non-zero temperatures and larger magnetic fields, we show that the magnetoconductivity can become non-analytic function of magnetic field proportional to $ B^{3/2}$. The non-analyticity is due to the singularity of the Berry curvature at small momenta, and we note that this contribution does not come from the Fermi surface.
Similar non-zero temperature, non-analytic magnetic field dependent magnetoconductivity was found in graphene\cite{Alekseev} and in updoped Weyl and Dirac semimetals. \cite{Janina} 

We also compute the magnetic field dependence of electric currents driven by temperature gradient. 
In which case the magnetoconductivity has quantitatively similar behavior as in the case of electric field driven currents.

We further include a time-reversal symmetry breaking tilt of the Dirac cones. 
We show that the tilt results in signatures of one dimensional chiral anomaly. 
Due to that there are two different in nature magnetic field dependent contributions to electric current. 
The first contribution to the current flows in the direction of the tilt, while second flows in the direction of the magnetic field. 
Both contributions are linear in magnetic field. 
The tilts of the Dirac cones have recently received theoretical attention. \cite{VolovikZubkov, Bergholtz_tilt, Soluyanov}
Various transport properties of Weyl semimetals due to tilt were studied in Refs. [\onlinecite{ZT, Udagawa_Bergholtz}].  
Systems with tilted Dirac cones were recently realized in Refs. [\onlinecite{tilt_exp1,tilt_exp2}].  
Present paper proposes linear magnetoconductivity as a distinct feature of time-reversal symmetry broken (e.g. ferromagnetic) Weyl semimetals, 
which are awaiting their experimental discovery.

\textit{Kinetic equation for Weyl semimetal}.
We consider a model of three dimensional Weyl fermions, which consists of two valleys with corresponding $s=\pm $ chiralities .
The Hamiltonian for $s=\pm$ valley is 
\begin{align}\label{ham}
H_{s} = C_{s}k^{(s)}_{z}  + s v \left[ \sigma_{x} k_{x} + \sigma_{y} k_{y} + \sigma_{z} k^{(s)}_{z}  \right] - \mu,
\end{align}
where $\sigma_{i}$ are $i=x,y,z$ Pauli matrices corresponding to a generalized spin degree of freedom. 
The valleys $s=\pm$ are split in momentum as $k_{z}^{(s)} = k_{z} + sQ$, where $Q$ is the splitting. 
However, as further calculations show, it is safe to ignore $Q$ in the derivations by shifting center of coordinates to $k_{z}^{(s)} = 0$ when studying corresponding valley.   
Diagonalization of the Hamiltonian gives the spectrum of the fermions,
$
\varepsilon^{(s)}_{{\bf k}\eta} = C_{s}k_{z}  +\eta  vk - \mu,
$
and the band velocity is obtained to be
$
{\tilde {\bf v}}^{(s)} = C_{s} {\bf e}_{z} + \eta v \frac{{\bf k}}{ k }.
$
Here $\mu$ is a Fermi energy, $v$ is a velocity, $\eta=\pm$ denote the conduction and valence bands correspondingly. $C_{s}$ is a parameter denoting the tilt of the Dirac cones. We choose it to be antisymmetric in valleys, $C_{+} = - C_{-}$. We assume that $v \gg \vert C\vert$, such that the Dirac cones are elliptical due to tilt. In the following we assume $\mu > 0$, such that all excitations occur in  conduction, $\eta = +$, band.  We focus only on the $\eta = +$ band in the following and omit $\eta$ symbol altogehter until mentioned.
  
In order to study the magnetotransport propeties of the system, we utilize the kinetic equation for the distribution function $n^{(s)}_{\bf k}$ of wave-packets constructed out of conduction bands fermions,
\begin{align}\label{kinetic_eq}
\frac{\partial n^{(s)}_{{\bf k}}}{\partial t}
+ {\dot {\bf k}}^{(s)} \frac{\partial n^{(s)}_{{\bf k}}}{\partial {\bf k}}
+ {\dot {\bf r}}^{(s)} \frac{\partial n^{(s)}_{{\bf k}}}{\partial {\bf r}}
= I_{\mathrm{coll}}\left[ n^{(s)}_{{\bf k}} \right],
\end{align}
where $I_{\mathrm{coll}}$ is the collision integral. Semiclassical equations of motion of wave packets in presence of electric ${\bf E}$ and magnetic ${\bf B}$ fields are, for example see Ref. [\onlinecite{Niu_rmp,Marder}] for a review,
\begin{align}
&
{\dot {\bf r}}^{(s)} = \frac{\partial \epsilon^{(s)}_{{\bf k}} }{\partial {\bf k}}
+{\dot {\bf k}}^{(s)}\times {\bf \Omega}^{(s)}_{{\bf k}\eta},
\nonumber
\\
&
 {\dot {\bf k}}^{(s)} = e{\bf E} + \frac{e}{c}{\dot {\bf r}}^{(s)}\times{\bf B},
\end{align}
where ${\bf \Omega}^{(s)}_{{\bf k}}$ is the Berry curvature and
where the energy of the wave-packet $\epsilon^{(s)}_{{\bf k}} = \varepsilon^{(s)}_{{\bf k}} - {\bf m}^{(s)}_{{\bf k}}{\bf B}$ is updated by orbital magnetization ${\bf m}^{(s)}_{{\bf k}}$ of the wave-packet.
Non-trivial Berry curvature and magnetization are consequences of a band degeneracy in the Brillouin zone.  
For example, in Weyl semimetal there is a linear touching of conduction and valence band at the Weyl points. 
In this case, one can show that the Berry curvature is a ${\bf k}-$ space analog of magnetic field created by a magnetic monopole located at these points, ${\bf \Omega}^{(s)}_{{\bf k}} = -s \frac{ {\bf k}}{2 k^3}$. 
The orbital magnetization is derived to be ${\bf m}^{(s)}_{{\bf k}} = -se\frac{v{\bf k}}{2ck^2}$. 
The Berry curvature and magnetization do not get affected by the tilt as the tilt enters the Hamiltonian with an identity matrix. 
One can spot an identity ${\bf m}^{(s)}_{\bf k} = \frac{e}{c}vk {\bm \Omega}^{(s)}_{\bf k}$.
After straightforward transformations in a system of equations on ${\dot {\bf r}}$ and ${\dot {\bf k}}$, one gets
\begin{align}\label{semiclassical}
& {\dot {\bf r}}^{(s)} = \frac{1}{\Delta^{(s)}_{{\bf k}}}
\left[ {\bf v}^{(s)} + e {\bf E}\times{\bf \Omega}^{(s)}_{{\bf k}} + \frac{e}{c}({\bf \Omega}^{(s)}_{{\bf k}} {\bf v}^{(s)} ){\bf B} \right]
\nonumber
\\
&
 {\dot {\bf k}}^{(s)} = \frac{1}{\Delta^{(s)}_{{\bf k}}}
\left[e{\bf E} + \frac{e}{c}{\bf v}^{(s)}\times {\bf B} + \frac{e^2}{c}\left({\bf E} {\bf B} \right){\bf \Omega}^{(s)}_{{\bf k}} \right],
\end{align}
where we defined  $\Delta^{(s)}_{{\bf k}} = 1 + \frac{e}{c}({\bf B} {\bf \Omega}^{(s)}_{{\bf k}})$, and velocity is ${\bf v}^{(s)} = \frac{\partial \epsilon^{(s)}_{{\bf k}}}{\partial \bf k}$, calculated to be  ${\bf v}^{(s)} = v\frac{{\bf k}}{k} [ 1+ \frac{2e}{c} ({\bf B}{\bf \Omega}^{(s)}_{{\bf k}} ) ] + s \frac{ev}{2c k^2}{\bf B} + s C{\bf e}_{z}$. We note that the semi-classical equations of motion are obtained from quantum kinetic equation by expanding the latter in small parameter $\frac{\omega_{\mathrm{c}}}{\mu} < 1$, where $\omega_{\mathrm{c}} = v\frac{eB}{ck_{\mathrm{F}}}$ with $k_{\mathrm{F}}$ being the Fermi momentum. \cite{Hong2000, SonYamamoto} When $k < \sqrt{\frac{eB}{c}}$ the Landau quantization of the energy levels must be made. 
This is a completely different problem and is out of scope of present paper.

The collision integral is assumed to only contain electron-impurity scattering.
The system of our study consists of two valleys, hence in general there are two distinct electron-impurity scattering processes, namely inter-valley and intra-valley.
Both can be conveniently captured by collision integral with Berry curvature modified density of states,\cite{Niu_rmp}
\begin{align}
I_{\mathrm{coll}}\left[ n^{(s)}_{{\bf k}} \right] = 
- \int_{{\bf k}^{\prime}} \Delta_{{\bf k}^{\prime}}^{(s^\prime)} \omega^{(ss^\prime)}_{{\bf k}^\prime,{\bf k}}  
\left[ n^{(s)}_{{\bf k}} - n^{(s^\prime)}_{{\bf k}^{\prime}} \right]
\delta( \epsilon^{(s)}_{{\bf k}} - \epsilon^{(s^\prime)}_{{\bf k}^{\prime}}),
\end{align}
where $\omega^{(ss^\prime)}_{{\bf k}^\prime,{\bf k}}$ is the probability of electron-impurity scattering between $s$ and $s^\prime$ valleys.
For short range impurities we obtain the collision integral, see a comment in Ref. [\onlinecite{PYM}],
\begin{align}
I_{\mathrm{coll}}[ n^{(\pm)}_{{\bf k}} ]
= \frac{{\bar n}^{(\pm)} - n^{(\pm)}_{{\bf k}}}{\tau}
+ \frac{{\bar n}^{(\mp)} - n^{(\pm)}_{{\bf k}}}{\tau_{\mathrm{V}}},
\end{align}
where ${\bar n}^{(s)} = \langle \Delta_{\bf k}^{(s)} n^{(s)}_{\bf k} \rangle$, here we defined $\langle .. \rangle =\frac{1}{4\pi} \int \sin(\theta)d\theta d\phi (..)$, $\tau$ is the intra-valley scattering time, and $\tau_{\mathrm{V}}$ is the inter-valley scattering time.
We assume that the scattering time is isotropic, i.e. angle indepenedent. It is a valid assumption for the $v \gg \vert C \vert $ case. It is convenient to rewrite the collision integral as
\begin{align}\label{collision}
I_{\mathrm{coll}}[ n^{(s)}_{{\bf k}} ] = \frac{1}{\tau^{*}} \left[ {\bar n}^{(s)} - n^{(s)}_{{\bf k}} \right] + \Lambda^{(s)} ,
\end{align}
where we introduced an important quantity
\begin{align}\label{lambdacollision}
\Lambda^{(\pm)} = \frac{1}{\tau_{\mathrm{V}}} \left[  {\bar n}^{(\mp)} - {\bar n}^{(\pm)} \right],
\end{align}
and $\tau^{*} =  \frac{\tau \tau_{\mathrm{V}}}{\tau+ \tau_{\mathrm{V}}}$ is the total scattering time.

Let us now approximate the kinetic equation. 
Assume that due to scattering life time, the distribution function is static, i.e. there is a steady state in the sysytem. 
That allows to drop time derivatives in kinetic equation. 
Multiply the rest of the kinetic equation for the $s=\pm$ valley by $\Delta_{\bf k}^{(s)}$ quantitiy and average over the angles. After that the $\propto \frac{1}{\tau^{*}}$ term in the collision integral (\ref{collision}) vanishes. Remaining equation reads as 
\begin{align}
\Lambda^{(s)} = \langle  \Delta_{\bf k}^{(s)}{\dot {\bf k}}^{(s)} \frac{\partial n_{{\bf k}}^{(s)} }{\partial {\bf k}} \rangle 
+ \langle \Delta_{\bf k}^{(s)}{\dot {\bf r}}^{(s)} \frac{\partial n_{{\bf k}}^{(s)}}{\partial {\bf r}} \rangle,
\end{align}
which allows us to find expression for $\Lambda^{(s)}$. To extract Berry curvature contribution to the magnetoconductivity, it is enough to use $\frac{\partial }{\partial {\bf k}} = \frac{\partial \epsilon_{{\bf k}}}{\partial {\bf k}}\frac{\partial }{\partial \epsilon_{{\bf k}}} = {\bf v}\frac{\partial }{\partial \epsilon_{{\bf k}}}$ in the expression for $\Lambda^{(s)}$. 
We next assume a gradient of temperature, which gives under an approximation of linear response a  
$
\frac{\partial n^{(s)}_{\bf k}}{\partial {\bf r}} = - \left[ {\bm \nabla} T({\bf r}) \right] \frac{\epsilon^{(s)}_{{\bf k}} }{T} \frac{\partial f^{(s)} }{\partial \epsilon^{(s)}_{{\bf k}}}
$
in the expression for the kinetic equation,
where $f^{(s)} = (e^{\epsilon^{(s)}_{{\bf k}}/T} + 1)^{-1}$ is an equilibrium Fermi-Dirac distribution function at, in general, non-zero temperature. 
Expanding in small parameters $\frac{\omega_{\mathrm{c}}}{\mu} < 1$ and $\frac{C}{v} < 1$, we get for $\Lambda^{(s)}$ an expression $\Lambda^{(s)} = \Lambda^{(s)}_{\mathrm{E}} + \Lambda^{(s)}_{\mathrm{T}}$ where, 
\begin{widetext}
\begin{align}
&\Lambda^{(s)}_{\mathrm{E}} = - s  \frac{ e^2 v^2  }{6ck}\left( {\bf E}{\bf B}\right)
\left( \frac{2}{vk}\frac{\partial f}{\partial \epsilon_{\bf k}} - \frac{\partial^2 f}{\partial \epsilon_{\bf k}^2}\right)
- s \frac{e}{2k}\frac{\left( {\bf E}{\bf C}\right)}{\vert C\vert}\left( 1 - \frac{v^2}{C^2} + \frac{v\mu}{C^2 k}\right)
\Theta\left( k - \frac{\mu}{v+\vert C \vert}\right)\Theta\left( \frac{\mu}{v- \vert C \vert} - k\right),
\label{lambdaE}
\\
&
\Lambda^{(s)}_{\mathrm{T}} = - s  \frac{ev}{6 k^2 Tc} \left( {\bf B} {\bm \nabla}T \right)
\left(  -2\epsilon_{\bf k} \frac{\partial f}{\partial \epsilon_{\bf k}}   + vk\frac{\partial f}{\partial \epsilon_{\bf k}} + vk\epsilon_{\bf k}\frac{\partial^2 f}{\partial \epsilon_{\bf k}^2}  \right)
-\frac{s}{3T}\left( {\bf C} {\bm \nabla}T \right)\left(vk\epsilon_{\bf k}\frac{\partial^2 f}{\partial \epsilon_{\bf k}^2} + vk\frac{\partial f}{\partial \epsilon_{\bf k}}
+3\epsilon_{\bf k}\frac{\partial f}{\partial \epsilon_{\bf k}}   \right),
\label{lambdaT}
\end{align}
\end{widetext}
where $\epsilon_{\bf k} = vk - \mu$ is used. 
We first observe that in the equations above $\Lambda^{(+)} = - \Lambda^{(-)}$, and it is consistent with definition of $\Lambda^{(s)}$ through collision integral, Eq. (\ref{lambdacollision}).
In case when $\tau_{\mathrm{V}} = \infty$, there is no steady state and the distribution function is a function of time. Then, after integrating the kinetic equation over $k$ we obtain $\frac{\partial N^{(s)}}{\partial t} + {\bm \nabla} {\bf J}^{(s)} = \int [ \Lambda^{(s)}_{\mathrm{E}} + \Lambda^{(s)}_{\mathrm{T}} ]\frac{ k^2 dk}{2\pi^2}$, where $N^{(s)} = \int {\bar n}^{(s)}\frac{k^2 dk}{2\pi^2}$ and is ${\bf J}^{(s)}$ is the local current.  
Calculations show $ \int [ \Lambda^{(s)}_{\mathrm{E}} + \Lambda^{(s)}_{\mathrm{T}} ]\frac{ k^2 dk}{2\pi^2} = s\frac{e^2}{4\pi^2 c}\left( {\bf E}{\bf B}\right) + s\frac{e }{4 \pi^2 c}g\left(-\frac{\mu}{T} \right)\left( {\bf B} {\bm \nabla}T \right) $, where $g(x) =\frac{xe^{x}}{e^{x}+1} - \ln\left[ 1+ e^{x}\right]$.
This is the chiral anomaly, \cite{Adler, BellJackiw, NN} i.e. non-conservation of chiral charge $N^{(+)} - N^{(-)}$ in presence of electric field or temperature gradient and magnetic field. We note that the tilt alone does not result in the chiral anomaly, but its presence in Eqs. (\ref{lambdaE}) and (\ref{lambdaT}), before momentum integraion, is due to one-dimensional chiral anomaly.
When $\tau_{\mathrm{V}} \neq \infty$ and there is a steady state, $\Lambda^{(s)}$ defined in Eq. (\ref{lambdacollision}) via the collision integral compensates the chiral anomaly given by Eqs. (\ref{lambdaE}) and (\ref{lambdaT}). Hence, we have the aforementioned $\Lambda^{(s)} = \Lambda^{(s)}_{\mathrm{E}}+\Lambda^{(s)}_{\mathrm{T}}$ equality.  

Since $\Lambda^{(s)}$ in steady state is defined by Eqs. (\ref{lambdaE}) and (\ref{lambdaT}), electron current will obtain magnetic field dependence through mechanism of chiral anomaly. To show this, we write an expression for the total electron current with subtracted magnetization parts, see \cite{Niu_rmp} for example,
\begin{align}\label{current}
{\bf j} 
&= e\sum_{s=\pm} \int_{{\bf k}}  \Delta_{\bf k}^{(s)}{\dot {\bf r}}^{(s)}n^{(s)}_{{\bf k}} 
\\
&+ e{\bm \nabla}\times T \sum_{s=\pm , \eta=\pm}
\int_{{\bf k}}{\bm \Omega}_{{\bf k}\eta}^{(s)}
\ln\left[ 1+e^{-\epsilon^{(s)}_{{\bf k}\eta}/T }\right], 
\nonumber
\end{align}
where for the sake of generality we restored the index $\eta$.  
The second term in the current corresponds to anomalous Hall effect \cite{BB,Niu_rmp,ZB}. 
It is very well studied, and we omit it in the following.
In order to calculate the magnetoconductivity, we rewrite the kinetic equation as
$ n^{(s)}_{{\bf k}}   
={\bar n^{(s)}} + \tau^{*} [   \Lambda^{(s)}
 - {\dot {\bf r}}^{(s)} \frac{\partial n_{{\bf k}}^{(s)}}{\partial {\bf r}} 
- {\dot {\bf k}}^{(s)} \frac{\partial n_{{\bf k}}^{(s)}}{\partial {\bf k}} ],$
keep in the right hand side only the terms linear in electric field and thermal gradient, and expand them in magnetic field assuming $\frac{\omega_{\mathrm{c}}}{\mu} <1$ and $\omega_{\mathrm{c}} \tau^{*}<1$. The latter condition is used in Zener-Jones method \cite{Zener_Jones, Ziman, PalMaslov} with which one can extract corrections to distribution function due to Lorentz force. They are very well studied and we omit them in the following.
In order to only extract the chiral anomaly contribution to the current in the limit $\tau_{\mathrm{V}} \gg \tau^{*}$ valid for large $Q$ - momentum splitting of $s=\pm$ valleys, it is enough to pick $n^{(s)}_{{\bf k}} = {\bar n^{(s)}}$. The expression for the magnetic field dependent contribution to the current due to chiral anomaly is then
\begin{align}
\delta{\bf j}_{\mathrm{\Lambda}} 
&
= e \sum_{s=\pm} \int_{{\bf k}} \left[ {\bf v}^{(s)} + \frac{e}{c}\left({\bm \Omega}_{\bf k}^{(s)}{\bf v}^{(s)} \right) {\bf B}\right]
 {\bar n}^{(s)}
\\
&
=
- \frac{2 e^2 v}{c} \tau_{\mathrm{V}} \int_{\bf k} \frac{{\bf k}}{k} \left({\bf B}{\bf \Omega}^{(+)}_{{\bf k}} \right)\Lambda^{(+)}
-  e\tau_{\mathrm{V}} {\bf C} \int_{\bf k} \Lambda^{(+)}.
\nonumber
\end{align}
In the following we list results for the magnetoconductivity as a sum $\delta {\bf j}_{\Lambda} = \delta {\bf j}^{[\mathrm{E}]}_{\mathrm{B}}+\delta {\bf j}^{[\mathrm{E}]}_{\mathrm{C}}+\delta {\bf j}^{[\mathrm{T}]}_{\mathrm{B}}+\delta {\bf j}^{[\mathrm{T}]}_{\mathrm{C}}$.

\textit{Quadratic and non-analytic magnetoconductivity}.
In this subsection we are going to study an un-tilted Dirac system, i.e. $C = 0$. 
To the lowest order in parameter $\frac{\omega_{\mathrm{c}}}{\mu}<1$ we get a contribution to the current, 
\begin{align}\label{result1}
\delta {\bf j}^{[\mathrm{E}]}_{\mathrm{B}}=
\tau_{\mathrm{V}} 
\frac{e^4 v^3}{36 \pi^2 c^2 \mu^2} \left( 1 + \frac{2\mu^2}{Tv\alpha} e^{-\mu/T} \right) \left({\bf E}{\bf B} \right){\bf B},
\end{align}
where $\alpha = \mathrm{max}\left( \ell^{-1}, L^{-1},\sqrt{\frac{e B}{c}}\right)$ is low-$k$ cut-off whose consequences are to be discussed below. 
There $\ell$ is the mean free path of an electron, $L$ is the system size, and we assumed $\mu \gg v\alpha$.
Obtained magnetoconductivity due to chiral anomaly qualitatively agrees with original results of Ref. [\onlinecite{SonSpivak,SpivakAndreev}]. 
Importantly, it has a positive sign.
Due to the small-$k$ cut-off we observe that when $T > T^{*} \approx \mu/ \ln\left[ \frac{k_{F}}{\sqrt{eB /c}} \right]$ given $\sqrt{\frac{e B}{c}}$ is the largest cut-off, the magnetic field dependent correction to the current written in (\ref{result1}) is 
\begin{align}\label{result1na}
\delta {\bf j}^{[\mathrm{E}]}_{\mathrm{B}} = 
 \tau_{\mathrm{V}}\frac{e^4 v^2}{18 \pi^2 c^2 T} e^{-\mu/T}\frac{\left({\bf E}{\bf B} \right){\bf B}}{\sqrt{eB/c}},
\end{align}
which is now proportional to $3/2$ power of the magnetic field, i.e. is a non-analyitc correction to the current. 
This contribution to the current is not from the Fermi surface but rather from the $k\sim 0$ region, where Berry curvature is singular.
As we show, regime of this result is magnetic field and impurity concentration dependent. 
Namely, in Weyl semimetals $\ell^{-1} = N\mu^2$, where $N$ is a parameter characterizing impurity concentration. 
We then write for the existence of the regime a condition $N\mu^2 < \ell_{\mathrm{B}}^{-1} \ll k_{\mathrm{F}}$. 
The latter inequality is satisfied by both Ioffe-Regel condition of a good metal, $k_{\mathrm{F}}\ell > 1$, and by the assumption of small magnetic fields $\frac{\omega_{\mathrm{c}}}{\mu} < 1$. 
The former inequality can be achieved by either decreasing parameter $N$ or increasing the magnetic field, while making sure the $\frac{\omega_{\mathrm{c}}}{\mu} < 1$ condition is always satisfied.

The magnetic field dependence of the current driven by temperature gradient (thermoelectric effect) due to chiral anomaly is
\begin{align}\label{result2}
\delta {\bf j}_{\mathrm{B}}^{[\mathrm{T}]}= \tau_{\mathrm{V}} \frac{e^3 v^3  }{36 \pi^2 c^2 \mu } 
\left[ \frac{2\pi^2}{3} \left( \frac{T}{\mu}\right)^2 +  \frac{2 \mu^2 }{T(v\alpha)}e^{-\mu/T} \right]\frac{ ({\bf B}{\bm \nabla}T){\bf B}}{T}
\end{align}
Again, when $\sqrt{\frac{e B}{c}}$ is the largest cut-off, and when the temperature is $T > T^{*} \approx \mu/ \ln\left[ \frac{k_{F}}{\sqrt{eB /c}} \right]$
we get non-analytic magnetic field dependence of the current,
\begin{align}\label{result2na}
\delta {\bf j}^{[\mathrm{T}]}_{\mathrm{B}} =  \tau_{\mathrm{V}} \frac{e^3 v^2 \mu}{  18 \pi^2 c^2 T }  
 e^{-\mu/T} \frac{ ({\bf B}{\bm \nabla}T){\bf B}}{T\sqrt{e B/c}}.
\end{align}
Consistent with the obtained results for the magnetoconductivity.

Formally, the divergence of integrals in expressions for the currents (\ref{result1}) and (\ref{result2}) signals the unapplicability of the semiclassical equations of motion (\ref{semiclassical}) at small $k$, hence a need for a cut-off (see a discussion after Eq. (\ref{semiclassical})).

\textit{Linear magnetoconductivity}.
In all previous derivations, terms linear in magnetic field dropped out due to angle integration. 
Let us now add a tilt to the spectrum, $C_{s} \neq 0$, as defined in the Hamiltonian (\ref{ham}). 
The tilt and magnetic field dependent contribution to the current driven by electric field is
\begin{align}\label{tilt_E}
\delta{\bf j}_{\mathrm{C}}^{[\mathrm{E}]} = 
-\tau_{\mathrm{V}}  \frac{e^3}{ \pi^2 c} \left[ \frac{1}{6}\left( {\bf E} {\bf C}\right){\bf B}
 + \frac{1}{4} \left({\bf E}{\bf B} \right) {\bf C} \right],
\end{align}
whose direction is in the direction of the magnetic field, and along the direction of the tilt.
We generalized the direction of the tilt by $C{\bf e}_{z}\rightarrow {\bf C}$ replacement.
The electric current driven by temperature gradient has similar corrections, namely
\begin{align}\label{tilt_T}
\delta{\bf j}^{[\mathrm{T}]}_{\mathrm{C}} & 
=  \tau_{\mathrm{V}} \frac{e^2  }{ \pi^2 c }g\left(-\frac{\mu}{T}\right) \left[ \frac{1}{6}\left(  {\bf C} {\bm \nabla}T  \right){\bf B}
+\frac{1}{4}  \left({\bf B}{\bm \nabla}T  \right) {\bf C} \right],
\end{align}
where $g(x)$ is defined after Eq. [\ref{lambdaT}], and $g(-x) \approx -xe^{-x}$ for $x>>1$.

The tilt, as introduced in expression (\ref{ham}) breaks the time-reversal symmetry, therefore the linear magnetoconductivity is not prohibited by the Onsager relations for the conductivity. 
We note that in deriving results (\ref{tilt_E}) and (\ref{tilt_T}), we approximated the scattering time to be isotropic, which is a valid assumption for $v \gg \vert C \vert$. 
We note that only in case of antisymmetric tilt, $C_{s} = s C$, we can approximate the kinetic equation in analytic way, and only in this case $\Lambda^{(+)} = - \Lambda^{(-)}$ relation holds. This relation is consistent with the steady state approximation of the kinetic equation. 
If the tilt is such that $C_{+}\neq -C_{-}$, there will be no steady state in the system and the distribution function will be a function of time. 
Finally, we note that first terms in (\ref{tilt_E}) and (\ref{tilt_T}) are due to last two terms of expression (\ref{lambdaE}) and (\ref{lambdaT}) correspondingly. 
We called them as the terms which resulted from one-dimensional chiral anomaly. \cite{NN}

\textit{Conclusions}.
We have studied various magnetotransport properties of a Weyl semimetal system due to chiral anomaly. 
Apart from predicted in Ref. [\onlinecite{SonSpivak,SpivakAndreev,BurkovPRL}] quadratic magnetoconductivity, 
we find a $\propto B^{3/2}$ non-analytic contribution to the magnetoconductivity at finite temperatures, see Eqs. (\ref{result1}) and (\ref{result1na}). 
This is due to the singular nature of the Berry curvature in Weyl semimetals. 

We have shown that the tilt of the Dirac cones results in two linear in magnetic field contributions to the magnetoconductivity. 
First contribution has a direction in magnetic field, and the second one in direction of the tilt, see Eqs. (\ref{tilt_E}).  
The tilt of the Dirac cones is responsible for mechanism similar to one-dimensional chiral anomaly, 
however its contribution to the three-dimensional chiral anomaly vanishes. 
The tilt breaks the time-reversal symmetry, therefore, linear magnetoconductivity is not prohibited.

We have studied magnetic field dependence of the current driven by the temperature gradient (thermoelectric effect) in Weyl semimetal due to chiral anomaly.
Results are qualitately consistent with the ones obtained for magnetoconductivity, see Eqs. (\ref{result2}), (\ref{result2na}), and (\ref{tilt_T}).

In present work we have developed analytical treatment of the electron-impurity collision integral for the Weyl semimetal under assumption of short range impurities. We have shown that inter-valley scattering processes can not be ignored and are responsible for the magnetotransport in Weyl semimetals due to chiral anomaly. A number of papers, for example Refs. [\onlinecite{Sasaki,LLF,SGT,Cortijo}], ignore these processes by setting $\tau_{\mathrm{V}} = \infty$ in Eq. (\ref{collision}), while [\onlinecite{SonSpivak, SpivakAndreev, Yip}] do include these processes. 
Approach of the present paper qualitatively agrees only with the one in Ref. [\onlinecite{SpivakAndreev}].
Magnetic field dependent contributions to the currents obtained in Refs. [\onlinecite{Sasaki,LLF,SGT,Cortijo}] come from the $\propto \tau^{*}$ corrections to the distribution function (see a discussion after Eq. \ref{current}). 
These corrections were ignored in present paper because of their parametric smallness due to $\tau_{\mathrm{V}} \gg \tau^{*}$ assumption.

We note that linear in magnetic field contributions to the current of untilted Weyl semimetals were obtained in Ref. [\onlinecite{Cortijo}].
Their origin is in interplay of orbital magnetization, Lorentz force, and quadratic corrections to the Dirac spectrum.  
Papers [\onlinecite{LLF,SGT}] also studied magnetothermal properties of Weyl semimetal. 
The low $k-$ divergence of the integrals defining thermoelectric coefficients were ignored in them. 

\textit{Acknowledgements}.
The author would like to thank D.L. Maslov, with whom the work was started, for inspiring discussions and support. 
He also acknowledges useful discussions with Saurabh Maiti, Xaver Neumeyer, A.A. Zyuzin, and especially A.Yu. Zyuzin.    
VAZ acknowledges support from the DOE Early Award DE-SC0014189. 
He also acknowledges A.F. Ioffe Physical-Technical Institute for warm hospitality.

\end{document}